\documentclass[twocolumn]{openjournal} 

\usepackage{orcidlink} 
\usepackage{hyperref}
\usepackage{url}

\usepackage{longtable}
\usepackage{booktabs}
\usepackage{tabu}
\usepackage{graphicx}
\usepackage{multirow}
\usepackage{ulem}
\usepackage{color}
\usepackage{amsmath}

\usepackage{xcolor}

\shorttitle{Iax not Ia due to Premature Ignition}
\shortauthors{Michaelis}

\graphicspath{{./}{img/}}

\begin{document}

\title{The Single-Degenerate Channel Leads to Type Iax and Not Type Ia Supernovae due to Premature Ignition}

\author{Amir Michaelis\,\orcidlink{0000-0002-1361-9115}} 
\affiliation{Department of Physics, Technion - Israel Institute of Technology, Haifa, 3200003, Israel; amichaelis@campus.technion.ac.il; }

\author{Hagai B. Perets\,\orcidlink{0000-0002-5004-199X}}
\affiliation{Department of Physics, Technion - Israel Institute of Technology, Haifa, 3200003, Israel; hperets@technion.ac.il}

% \date{\today}

\begin{abstract}
 Type Ia supernovae (SNe Ia) are a critical tool for cosmology and galactic enrichment, yet the progenitor systems of normal SNe Ia remain a central puzzle. The long-debated single-degenerate (SD) channel, where a white dwarf (WD) accretes mass from a companion, faces major observational conflicts. Here, we present 3D hydrodynamic simulations that resolve these tensions by showing a fundamental dichotomy: accreting WDs predominantly ignite prematurely at sub-Chandrasekhar masses, producing low-energy, incomplete explosions consistent with Type Iax supernovae. Only WDs reaching a narrow mass threshold of 1.37 M$_\odot$ undergo complete destruction, characteristic of normal SNe Ia. This "safety valve" mechanism effectively recasts the SD channel as the main pathway to SNe Iax, not normal SNe Ia, providing a unified explanation for the observed scarcity of progenitor signatures in the latter and suggesting alternative channels dominate normal SNe Ia production.
\end{abstract}
   

% ==================================
\section{Introduction} 
\label{sec:intro}
% ==================================
Type Ia supernovae (SNe Ia) represent a fundamental class of stellar explosions with far-reaching implications for astrophysics. These thermonuclear events play crucial roles in galactic chemical evolution as primary producers of iron-peak elements \cite{Iwamoto1999, Seitenzahl2013, Palicio2023}, influence the energy budget and dynamics of their host galaxies \cite{Taylor2014, Liang2024, Palicio2024b}, and serve as laboratories for extreme physics under degenerate conditions \cite{Hillebrandt2013}. Despite their importance across multiple domains of astrophysics, the nature of their progenitors and explosion mechanisms remains one of the most persistent open questions in stellar evolution theory.

Traditionally, progenitor models have been divided broadly into single-degenerate (SD) scenarios, where a carbon-oxygen (CO) white dwarf (WD) accretes mass from a non-degenerate companion, and double-degenerate (DD) scenarios involving the merger of two WDs [e.g. see reviews by \cite{Maoz2014review, Soker2024review, Ruiter2025review}]. Both scenarios face theoretical challenges in explaining the observed homogeneity and diversity of SNe Ia properties, timing of explosions, and nucleosynthetic yields. In particular, the single-degenerate channel faces stringent empirical limits from the paucity of luminous supersoft X-ray sources (SSS) and from nova statistics (including fast recurrent novae; RNe), which together challenge a dominant SD contribution to normal SNe Ia \cite{Gilfanov2010Nature,Johansson2014MNRAS,Nielsen2014MNRAS,Soraisam2015AandA,Shafter2017ApJ}.

Additional tensions for SD–Ia include (i) stringent limits on stripped H/He in normal-Ia nebular spectra, (ii) non-detections of luminous surviving companions and of predicted early blue/UV “companion-shock” signals, (iii) deep radio/X-ray limits implying very low mass-loss rates, and (iv) potential rate and delay-time distribution mismatches \cite{Mattila2005,Lundqvist2015,Maguire2016,Botyanszki2018,Li2011,Kerzendorf2018,Kasen2010,Hayden2010,Bianco2011,Chomiuk2012,Chomiuk2016,Margutti2014,Maoz2014review}; see
Ref. \cite{Maoz2014review} for a review. 

Here we show that instead of the SD channel leading to normal type Ia SNe, they produce mostly (if not only) fainter type Iax SNe through premature ignition at lower masses than envisioned before, and partial deflagration. The various challenges and tensions with SD channel for normal type Ia SNe are then naturally resolved and become consistent with the properties of type Iax SNe. 

Recent observational advances have revealed that thermonuclear supernovae exhibit considerably more diversity than the classical SNe Ia classification initially suggested \cite{Tau17}. Modern transient surveys have identified numerous subclasses of thermonuclear events, including the extremely luminous "super-Chandrasekhar" events \cite{Howell2006, Scalzo2010, Phillips2024}, the peculiar 2005E-like calcium-rich gap transients \cite{per+10, Kasliwal2012}, the rapidly declining "SNe Ia-02es-like" objects \cite{Ganeshalingam2012, White2015, Cao2016, Bose2025}, and the spectroscopically distinct "SNe Iax" \cite{Li2003, Foley2013}. This diversity challenges our understanding of thermonuclear explosion mechanisms and the evolutionary pathways leading to them.

Among these peculiar transients, Type Iax supernovae (SNe Iax) constitute a prominent subclass, initially identified with SN 2002cx \cite{Li2003, Foley2013}. Unlike classical SNe Ia, SNe Iax exhibit lower peak luminosities, lower ejecta velocities (approximately $2,000-8,000$ km s$^{-1}$), asymmetric ejecta, and smaller $^{56}\text{Ni}$ yields ranging from $0.03$ to $0.2~M_{\odot}$ \cite{Foley2013, Jha2017}. Moreover, SNe Iax spectra show persistent low-ionization features at late phases, indicative of incomplete burning and suggesting a partially exploded remnant WD \cite{Sahu2008, Jha2006, Foley2016, Camacho-Neves2023, Schwab2025}.

Earlier theoretical work has demonstrated that partial deflagration explosions of near-Chandrasekhar-mass CO WDs can naturally produce observed SN Iax characteristics, including moderate ejecta velocities, modest nickel yields, and remnant natal kicks due to asymmetric ejection \cite{Jordan2012b, Kromer2013, Fink2014, Mehta2024}. Specifically, Jordan et al.\cite{Jordan2012b} suggested that incomplete deflagration of near-Chandrasekhar mass WDs leads to asymmetric ejecta, natal kicks to surviving WD remnants, and lower yields of radioactive $^{56}\text{Ni}$. Nevertheless, this scenario raises the critical unresolved question of why some CO WDs experience only partial deflagration, resulting in Type Iax, whereas others undergo complete deflagration, transitioning into detonation, producing normal SNe Ia. A related challenge in SD models involves the limitations on WD growth due to nova eruptions. As WDs accrete material from their companions, periodic nova explosions can expel accreted matter, potentially halting the WD growth towards the Chandrasekhar limit \cite{Iben1984, Yaron2005, Starrfield2020, Hillman2021, Hillman2025}.

Analogous to the nova-eruptions limiting challenge, we propose that a similar limiting phenomenon occurs not only via surface novae but also through internal deflagration ignition processes. In particular, CO WDs may ignite prematurely at sub-Chandrasekhar masses, initiating partial deflagrations that fail to transition to full detonation. Such early ignition, triggered by lower central densities, may preferentially yield Type Iax events rather than classical SNe Ia, and will expel significant mass, not allowing for further growth of the WD and precluding a full normal type Ia SNe from this channel. In this study, we employ detailed three-dimensional hydrodynamic simulations to explore the deflagration mechanism in single-degenerate sub-Chandrasekhar CO WDs. We systematically vary WD masses and ignition conditions to determine the circumstances under which partial deflagrations, rather than full detonations, occur. Our results demonstrate that ignition at lower WD masses consistently produces explosion signatures consistent with the observational properties of SNe Iax, thereby offering theoretical support for the hypothesis of premature ignition as a critical pathway to these peculiar explosions. This insight provides a coherent explanation for the diversity observed among thermonuclear supernovae and the population of sub-Chandrasekhar WDs that never reach full detonation but rather manifest observationally as SNe Iax. Our findings indicate that premature ignition at sub-Chandrasekhar masses imposes a significant constraint on the single-degenerate scenario, suggesting it rarely leads to typical Type Ia SNe.

% ==================================
\section{Results}
\label{sec:res}
\begin{figure}
	\centering
    \includegraphics[width=0.5\textwidth]{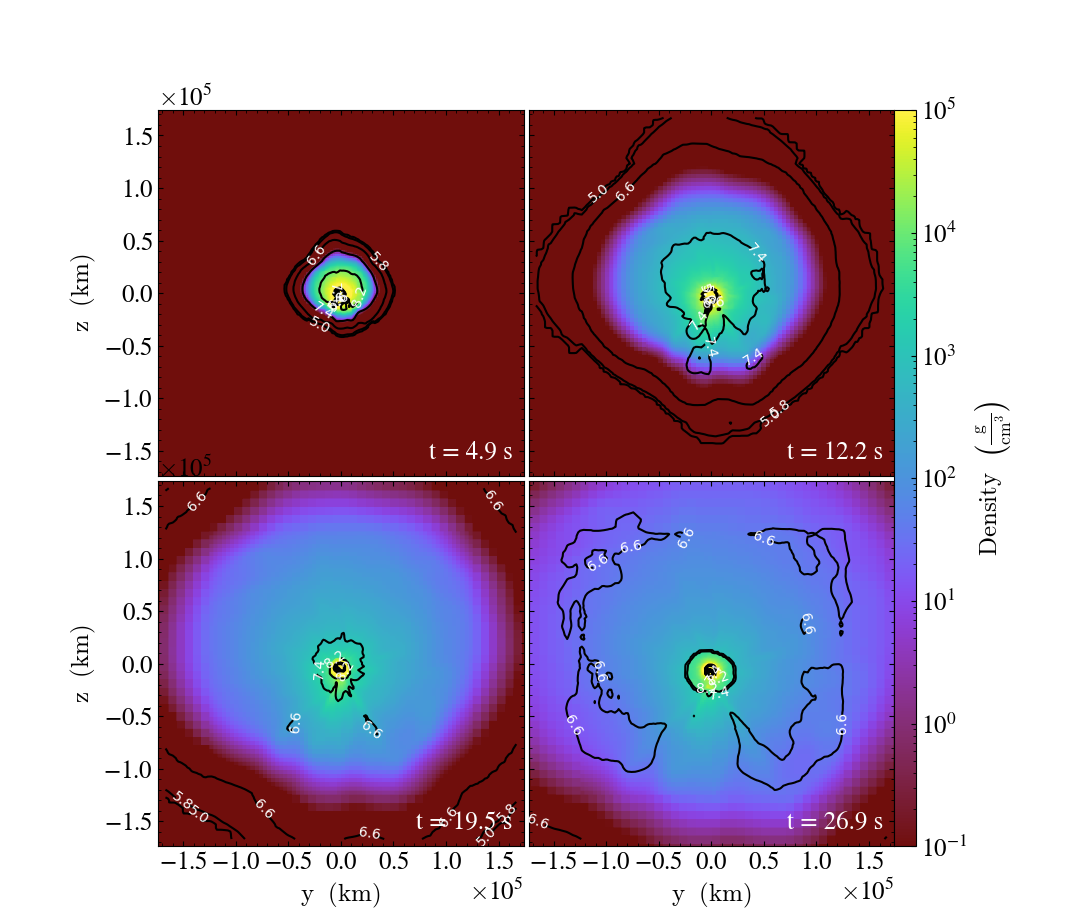}    
    \includegraphics[width=0.5\textwidth]{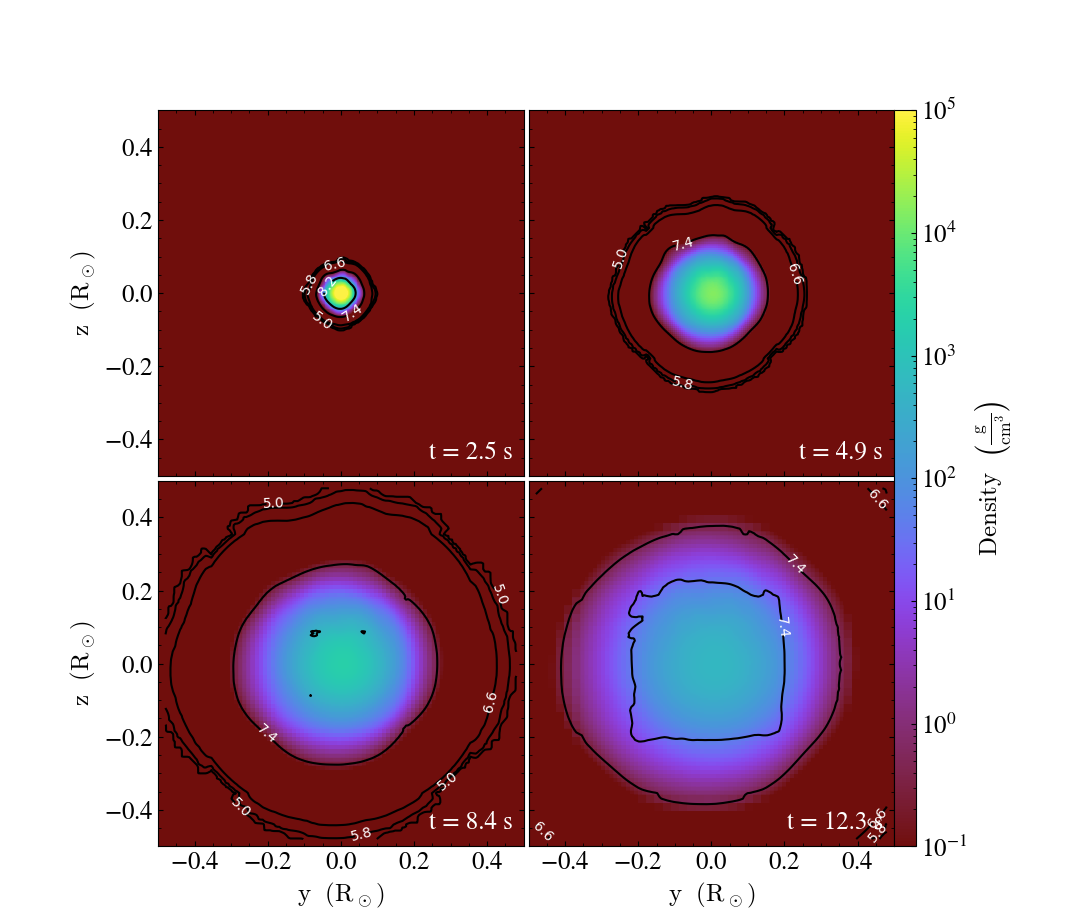}
    \caption{\textbf{Time evolution of the deflagration front and density structure for representative models.} \textbf{Top row:} $1.330~M_{\odot}$ WD (partial deflagration). Snapshots are shown at $2.5$ and $4.9$ seconds. \textbf{Bottom row:} $1.374~M_{\odot}$ WD (complete disruption). Snapshots are shown at $8.4$ and $12.3$ seconds. The color map changes from red ($0.1$ g cm$^{-3}$) up to yellow ($10^5$ g cm$^{-3}$). The contours represent the temperature at $5$ different values. The asymmetric burning in the partial deflagration case contrasts with the nearly symmetric energy release in the complete disruption case.}
    \label{fig:density_evolution}
\end{figure}

\begin{figure}
    \includegraphics[width=0.5\textwidth]{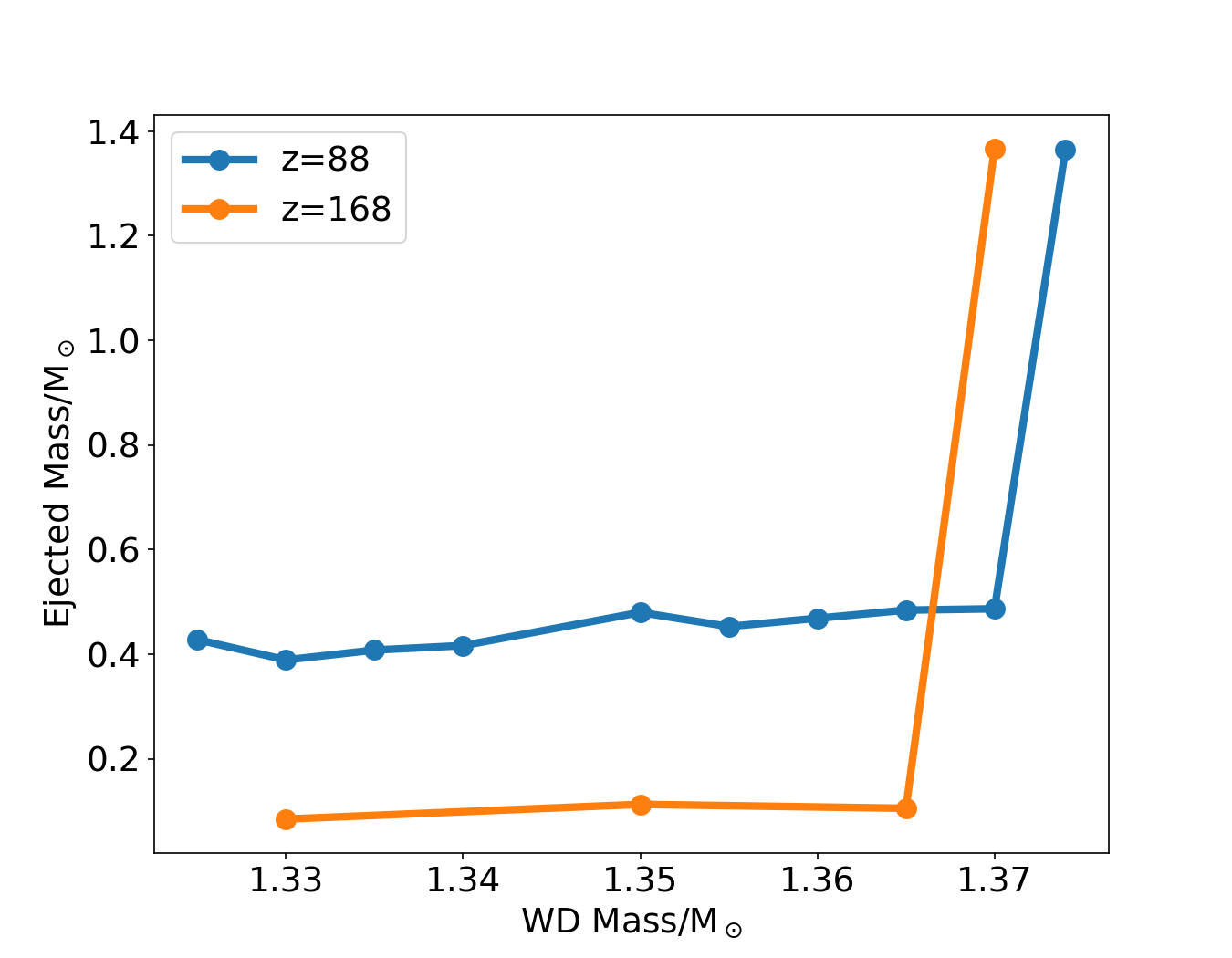}
    \caption{\textbf{Relationship between white dwarf mass and ejected mass}. The blue line represents models with an ignition offset of $88$ km, while the orange line represents models with an ignition offset of $168$ km. Note the sharp transition in ejected material at approximately $1.370~M_{\odot}$, corresponding to the boundary between partial deflagrations and complete disruptions. There is a strong dependency of the ejected material on the ignition offset; for a given offset, we see a much more subtle change in the ejected mass.}
    \label{fig:mass_dependence}
\end{figure}

\begin{figure}
    \includegraphics[width=0.5\textwidth]{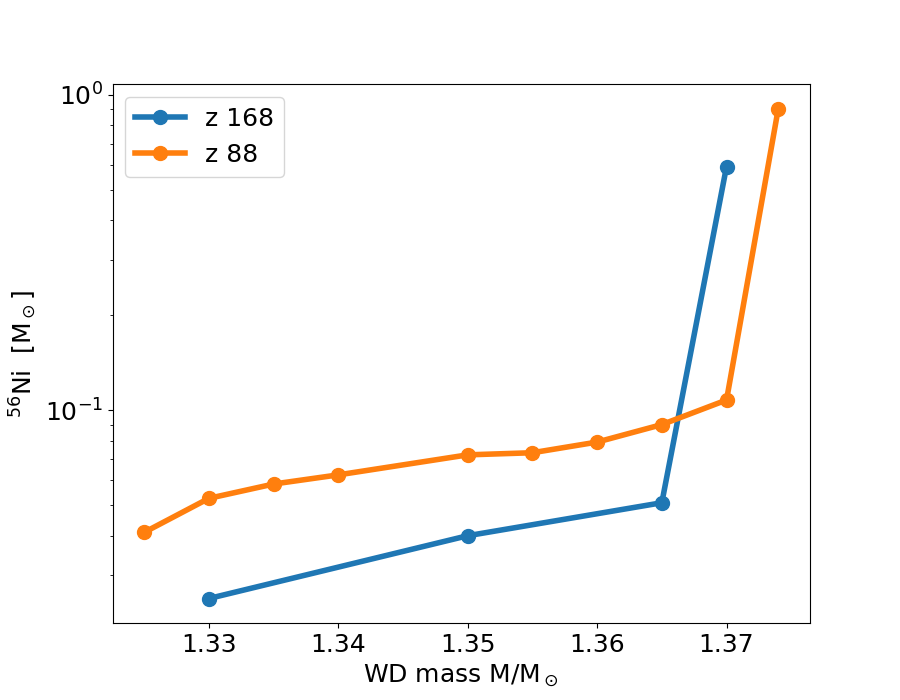}
    \caption{\textbf{Relationship between white dwarf mass and $^{56}\text{Ni}$ yield.} The blue line represents models with an ignition offset of $88$ km, while the orange line represents models with an ignition offset of $168$ km. Note the sharp transition in nickel production at approximately $1.370~M_{\odot}$, corresponding to the boundary between partial deflagrations and complete disruptions. A clear dependency between the nickel yield from the explosion and the initial WD mass is evident.}
    \label{fig:mass_vs_56ni}
\end{figure}

\begin{figure}
    \includegraphics[width=0.5\textwidth]{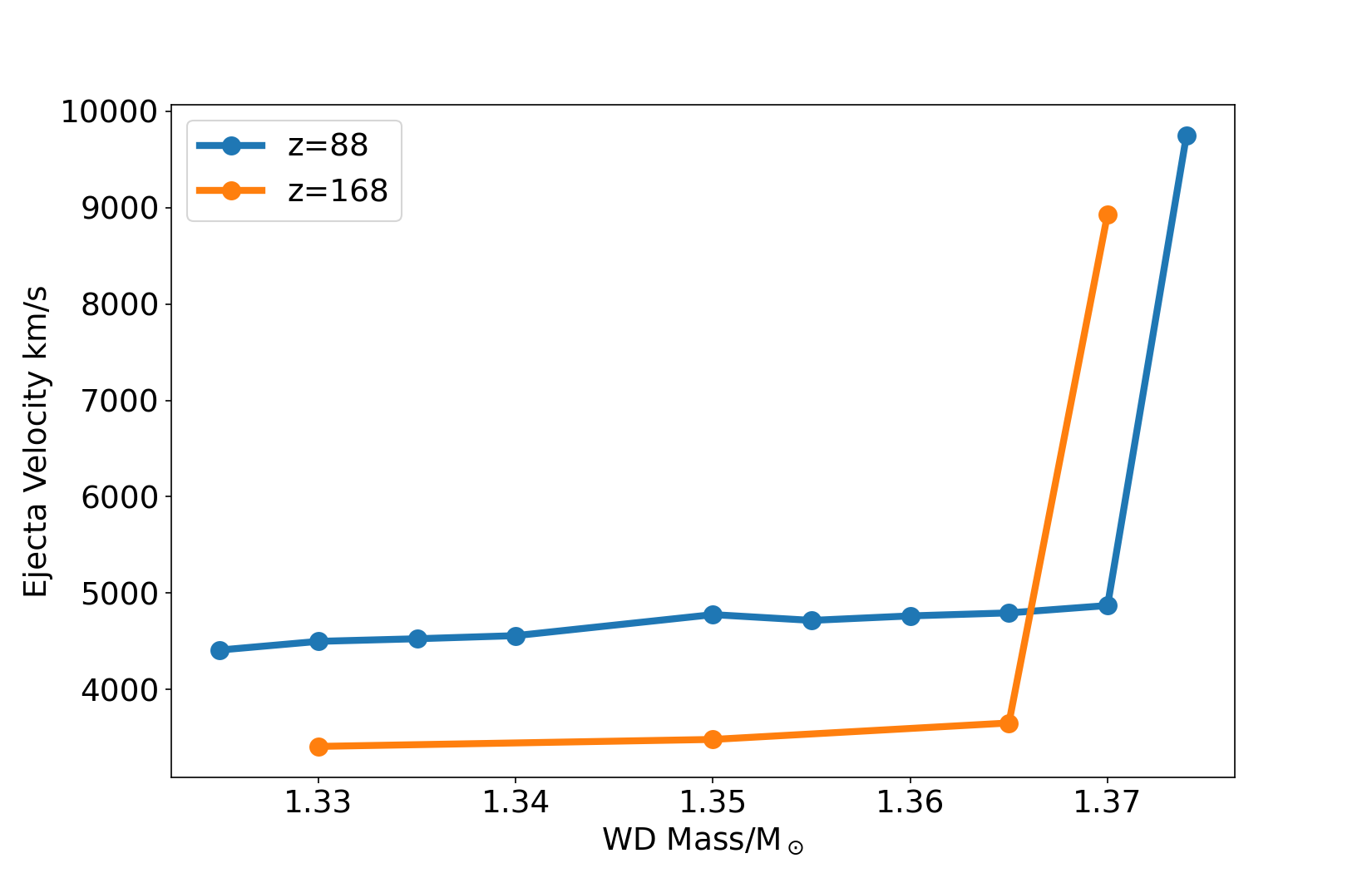}
    \caption{\textbf{Relationship between WD mass and mean ejecta velocity.} The blue line represents ignition points centered at $z=88$ km, and the yellow line represents the center at $z=168$ km. The destroyed WDs (models w1374-z88 and w1370-z168) show velocities typical of Type Ia supernovae ($\sim 10,000$ km s$^{-1}$). The partially exploded WDs have typical velocities of Type Iax supernovae, slowly ranging between $4000-5000$ km s$^{-1}$.}
    \label{fig:wd_vel_ucom}
\end{figure}

% ==================================
Our three-dimensional hydrodynamic simulations (see Fig. \ref{fig:density_evolution} and methods section) reveal a clear mass-dependent transition in the explosion outcomes of ignited white dwarfs. For WDs with masses $\lesssim 1.365~M_{\odot}$, we consistently observe partial deflagrations that fail to unbind the entire star. These incomplete explosions leave behind a remnant WD, typically ejecting $0.1-0.5~M_{\odot}$ of material with characteristic velocities of $3,000-5,000$ km s$^{-1}$ and producing modest amounts of $^{56}\text{Ni}$ ($0.03-0.09~M_{\odot}$). In contrast, WDs with masses $\gtrsim 1.370~M_{\odot}$ undergo complete disruption, with no remaining bound remnant (see Fig. \ref{fig:mass_dependence}). These energetic explosions produce significantly higher nickel yields ($0.5-1.2~M_{\odot}$; Fig. \ref{fig:mass_vs_56ni}) and eject material at much higher velocities ($8,000-10,000$ km s$^{-1}$; Fig. \ref{fig:wd_vel_ucom}). The transition between these two regimes occurs over a remarkably narrow mass range of approximately $0.005~M_{\odot}$. The relationship between the mass of the white dwarf and the unbounded mass ejected (Figure \ref{fig:mass_dependence}) and the $^{56}\text{Ni}$ yield (Figure \ref{fig:mass_vs_56ni}) clearly illustrate this sharp transition in nickel production at approximately $1.370~M_{\odot}$. The ejected mass and the velocity show a much more moderate dependency on the initial WD mass for partial deflagrations (Figure \ref{fig:wd_vel_ucom}).

Our simulations also demonstrate significant sensitivity to the ignition geometry, particularly the offset distance of the ignition region from the WD center. As shown in Table \ref{tab:models}, models with an ignition offset of $168$ km consistently produce weaker explosions compared to their counterparts with an ignition offset of $88$ km, for the same WD mass. For instance, model w1350-z168 ejects only $0.113~M_{\odot}$ with $0.040~M_{\odot}$ of $^{56}\text{Ni}$, whereas model w1350-z88 ejects $0.480~M_{\odot}$ with $0.072~M_{\odot}$ of $^{56}\text{Ni}$. This difference is particularly pronounced for intermediate-mass WDs ($1.350-1.365~M_{\odot}$), where the ignition offset can determine whether the explosion ejects a significant fraction of the WD mass or only a small outer layer. Interestingly, for the highest-mass WDs in our study ($1.370~M_{\odot}$ and above), the ignition offset primarily affects the energy distribution and nucleosynthetic yields rather than the overall explosion outcome. Both models w1370-z88 and w1370-z168 lead to complete disruption of the WD, although with different $^{56}\text{Ni}$ yields ($0.108$ vs. $0.593~M_{\odot}$). This suggests that while the ignition geometry modulates the explosion strength, the WD mass ultimately determines whether the star undergoes complete disruption.

\begin{figure}    
    \includegraphics[width=0.5\textwidth]{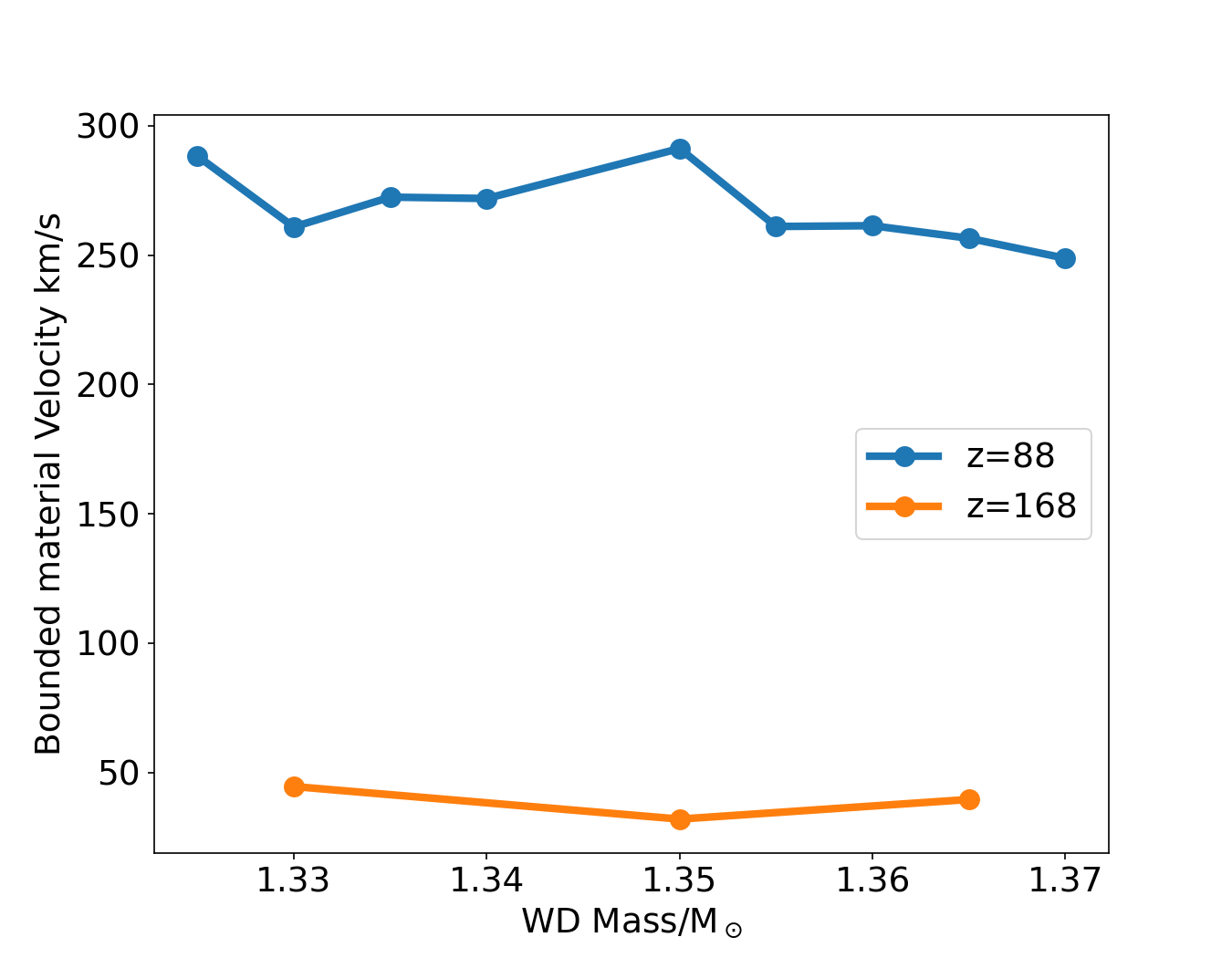}
    \caption{\textbf{Relationship between the WD mass and the kick velocity of the bounded material.} The blue line represents ignition points centered at $z=88$ km and the yellow line represents the center at $z=168$ km. The destroyed WDs (models w1374-z88 and w1370-z168) are omitted from this plot as there is no kick velocity in these cases.}
    \label{fig:wd_vel_bcom}
\end{figure}

The explosion dynamics of our simulated WDs exhibit distinctive characteristics based on mass. In lower-mass WDs, the deflagration front propagates asymmetrically through the star, burning only a fraction of the available fuel before the expansion of the star quenches the flame. This asymmetric burning results in asymmetric ejecta and imparts a natal kick to the remnant WD in the direction opposite to the primary ejecta. Our simulations yield kick velocities ranging from $100$ to $500$ km s$^{-1}$, with higher kicks generally associated with larger ejecta masses (see Figure \ref{fig:wd_vel_bcom}). The surviving remnants from partial deflagrations typically have masses of $0.9-1.2~M_{\odot}$ and significantly expanded radii compared to their preexplosion state. These remnants are heated by deflagration and show strong convective mixing, leaving them with a unique composition profile that includes partially burned material and ash from nuclear burning. In contrast, the highest-mass WDs in our study develop a more symmetric and energetic burning front that transitions to a detonation wave, consuming the entire star and leaving no bound remnant. The transition from deflagration to detonation occurs when the deflagration front reaches regions of critical density and temperature conditions, typically in the range of $1-2 \times 10^9$ g cm$^{-3}$. Figure \ref{fig:density_evolution} illustrates the time evolution of the deflagration front and density structure for representative models of each mass regime.

\begin{figure}
    \includegraphics[width=0.5\textwidth]{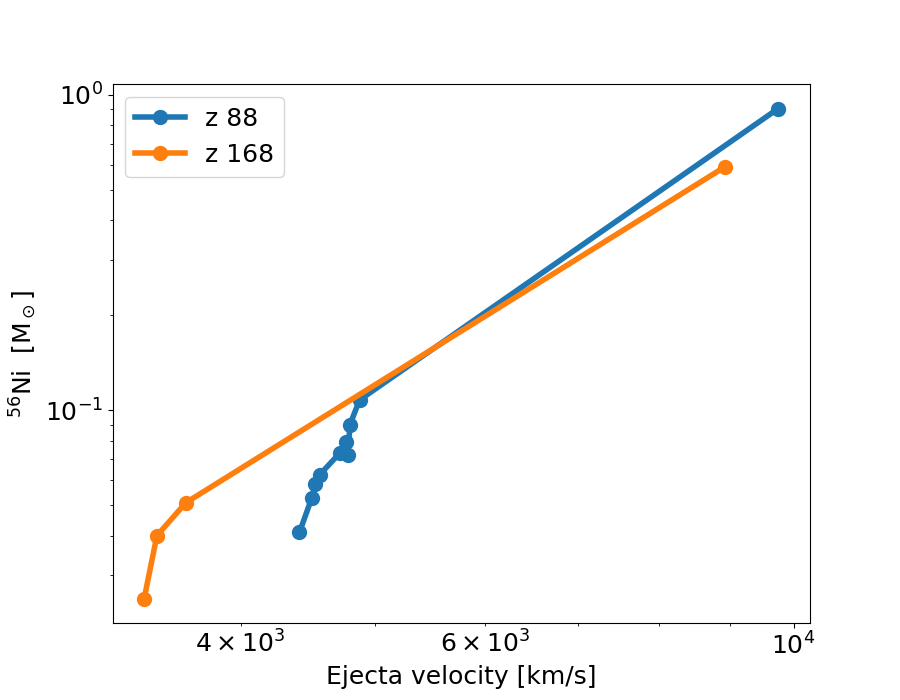}
    \caption{\textbf{$^{56}\text{Ni}$ ejecta for selected models.} A log-log plot of nickel in units of solar mass against the velocity of the nickel shell ejecta in units of km s$^{-1}$. The blue line represents ignition points centered at $z=88$ km, and the yellow line represents the center at $z=168$ km. The destroyed WDs (models w1374-z88 and w1370-z168) show nickel and velocities typical of Type Ia SN. The partially exploded WDs have typical nickel and velocities of type Iax SNe.}
    \label{fig:v_ni_1}
\end{figure}

The nucleosynthetic yields from our simulations provide further insights into the explosion mechanisms and their observational signatures. For partial deflagrations (lower-mass WDs), the composition of the ejecta is dominated by intermediate-mass elements (IMEs) such as silicon, sulfur, and calcium, with relatively small amounts of iron-group elements. The $^{56}\text{Ni}$ fraction in these models typically constitutes $10-20\%$ of the mass of the ejecta. This composition is consistent with the observed spectra of SNe Iax, which show strong IME features and relatively weak iron-group signatures at early times. In contrast, the complete disruption models (higher-mass WDs) produce ejecta rich in iron-group elements, with $^{56}\text{Ni}$ accounting for $40-60\%$ of the ejecta mass. This composition aligns well with normal SNe Ia, which exhibit strong iron-group element features in their spectra. The sharp transition in nucleosynthetic yields observed across our mass sequence provides a natural explanation for the distinct observational properties of SNe Iax versus normal SNe Ia, with the WD mass at ignition being the primary determining factor. Figure \ref{fig:v_ni_1} presents the $^{56}\text{Ni}$ yield against ejecta velocity for the different models.

% ==================================
\section{Discussion}
\label{sec:discussion}
\begin{figure}
    \includegraphics[width=0.5\textwidth]{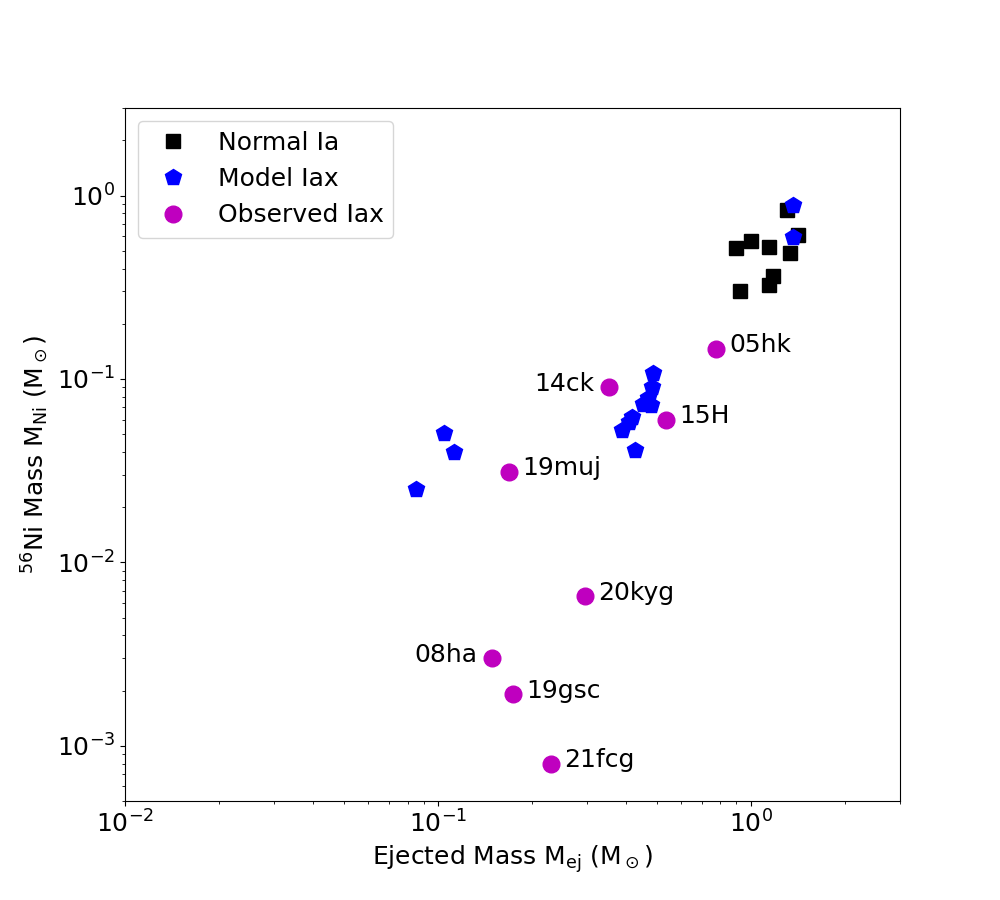}
    \caption{\textbf{Comparison between our simulation results and observed properties of SNe Iax.} Our simulation results (marker with blue pentagons) are compared with observed SNe Iax properties (mark with magenta circles). Normal type Ia (marked with black squares) are added for perspective. The vertical axis shows the $^{56}\text{Ni}$ mass yield, while the horizontal axis shows the ejected mass. The observed properties are from Srivastav et al. \cite{Srivastav2020} figure 16. Our partial deflagration models populate the same region of parameter space as observed SNe Iax, while our complete disruption models align with normal SNe Ia.}
    \label{fig:Ni_EjeMass_observations_comparison}
\end{figure}

% ==================================
Our simulation results offer a compelling explanation for the observed characteristics of Type Iax supernovae. The partial deflagrations of sub-Chandrasekhar mass WDs consistently produce explosions with ejecta velocities,
$^{56}\text{Ni}$ yields, and total ejected masses that align remarkably well with observed SNe Iax properties \cite{Foley2013, Jha2017, Srivastav2020}. Our simulated ejecta velocities ($3,000-5,000$ km s$^{-1}$) fall squarely within the range of observed photospheric velocities in SNe Iax ($2,000-8,000$ km s$^{-1}$). Similarly, our $^{56}\text{Ni}$ yields ($0.03-0.1~M_{\odot}$) match the inferred radioactive yields of observed events, which typically range from $0.03$ to $0.2~M_{\odot}$. Figure \ref{fig:Ni_EjeMass_observations_comparison} directly compares our model predictions with observed SNe Iax properties. The heterogeneity within the SNe Iax class can be naturally explained within our framework by variations in the WD mass and ignition conditions. We should note that 2008ha-like fainter SNe (e.g., \cite{Foley2009, Stritzinger2014}) were not produced by our models. It is possible that events like SN 2008ha and SN 2010ae, with inferred $^{56}\text{Ni}$ masses of only $\sim 0.003~M_{\odot}$, could result from ignitions occurring at even lower WD masses or more extreme offsets than those explored in our current parameter study. Alternatively, while 2008ha-like SNe were suggested to be an integral part of the Iax SNe \cite{Foley2013}, including both brighter 2002cx-like SNe and fainter 2008ha-like SNe, observations do show a clear luminosity gap between brighter and fainter Iax SNe, raising the possibility that they result from different channels. Hence, not obtaining very faint SNe in our models might be consistent with such a possibility of two distinct channels, and may support it. On the brighter side of Iax SNe, the relatively luminous SN 2012Z \cite{Yamanaka2015}, with an estimated $^{56}\text{Ni}$ mass of $\sim 0.18~M_{\odot}$, might represent an explosion of a WD near the upper mass boundary of the partial deflagration regime. The persistent detection of bound remnants in our simulations aligns with late-time observations of SNe Iax, which often show peculiar spectroscopic features inconsistent with complete disruption scenarios \cite{Jha2006, Foley2016}. These remnants could potentially be observed as unusual WDs with peculiar atmospheric compositions, high velocities, and inflated radii, and were possibly observed, e.g. Ref. \cite{ven+17}.

A key empirical objection to SD progenitors is the predicted prevalence of luminous supersoft X-ray sources (SSSs) and rapidly recurring novae (RNe) during the terminal growth to $M_{\rm Ch}$. Observations of early-type galaxies show soft X-ray fluxes $\sim30$–$50\times$ below naive SD expectations \cite{Gilfanov2010Nature}, with additional constraints from the non-detection of SSS-ionized He\,\textsc{ii} nebular emission \cite{Johansson2014MNRAS} and pre-explosion X-ray limits for nearby SNe Ia \cite{Nielsen2014MNRAS}. Nova statistics in M31 likewise limit the fraction of SNe Ia whose final $\sim0.1~M_\odot$ is accreted via unstable burning to at most a few percent \cite{Soraisam2015AandA}, consistent with Galactic nova-rate budgets \cite{Shafter2017ApJ}. In our framework, accreting CO WDs ignite prematurely at sub-Chandrasekhar masses and produce Iax-like partial deflagrations, which \emph{truncate} the high-mass steady-burning phase and remove the fastest high-mass RN cycles. A closed-form estimate (see Methods) shows that lowering the effective cap from $1.38$ to $1.32~M_\odot$ suppresses the SSS time budget by $\sim70$–$85\%$ for tracks entering stability near $1.30~M_\odot$, while removing $\sim40$–$50\%$ of the high-mass RN contribution; at $1.30~M_\odot$ the SSS exposure vanishes in this toy model. Thus, if the SD channel predominantly yields Iax, the SSS and RN tensions are naturally mitigated without appealing to extreme obscuration or unusually short duty cycles.

Recasting the single-degenerate channel as predominantly producing Iax naturally mitigates several independent tensions for SD–Ia. 
(i) \textit{Stripped H/He:} normal-Ia nebular spectra rarely show the $\gtrsim10^{-3}$–$10^{-2}\,M_\odot$ of stripped companion material expected for SD–Ia, whereas the weaker, slower Iax partial deflagrations suppress stripping, enhance fallback/mixing, and yield fainter, broader late-time lines consistent with non-detections \cite{Mattila2005,Lundqvist2015,Maguire2016,Botyanszki2018}. 
(ii) \textit{Surviving companions and early shocks:} pre- and post-explosion searches seldom find luminous non-degenerate companions for normal Ia, and early blue/UV “companion-shock” flashes are rare \cite{Li2011,Kerzendorf2018,Kasen2010,Hayden2010,Bianco2011}; in Iax, lower energies/velocities shift these signals below typical detection thresholds. 
(iii) \textit{Radio/X-ray circumstellar limits:} deep non-detections for nearby normal Ia imply very low mass-loss rates \cite{Chomiuk2012,Chomiuk2016,Margutti2014}, while intrinsically faint Iax high-energy emission and occasional He/CSM hints remain compatible with SD donors within the Iax framework. 
(iv) \textit{Hosts and delays:} Iax favor late-type, star-forming hosts and shorter delays (e.g. \cite{Lyman2013}), consistent with SD accretors expected from most population synthesis studies of this channel \cite{Maoz2014review}, whereas the full normal-Ia population spans older hosts and a $\sim t^{-1}$ delay-time distribution less naturally dominated by SD. 
Thus, a single mechanism—premature, sub-Chandrasekhar ignition—simultaneously alleviates the SSS/RNe expectations and these broader SD signatures.

Our finding that WDs with masses below $\sim 1.370~M_{\odot}$ consistently undergo partial rather than complete explosions has major implications for single-degenerate evolution scenarios. This result suggests that the majority of accreting WDs in single-degenerate systems may never reach the conditions necessary for normal Type Ia explosions. In traditional single-degenerate scenarios, a CO WD accretes hydrogen-rich material from a non-degenerate companion, growing toward the Chandrasekhar limit \cite{Nomoto1982, Hachisu1996}. However, this growth process faces several theoretical challenges, including nova eruptions that may expel more mass than is accreted \cite{Yaron2005, Hillman2016}. Our work introduces an additional, potentially more fundamental limitation: even if a WD successfully accretes enough material to reach masses of $1.32-1.37~M_{\odot}$, it becomes increasingly susceptible to premature ignition, leading to a partial deflagration. This premature ignition scenario resembles a "safety valve" in single-degenerate evolution: as the WD's mass increases and its central density rises, the probability of spontaneous carbon ignition grows dramatically. Once ignited, the partial deflagration ejects a substantial fraction of the accreted mass, effectively resetting the growth process. In principle, this cycle could repeat multiple times during a binary's evolution, with each partial explosion manifesting as a SN Iax, however, the original mass-growth towards a near-Chandrasekhar mass WD already exhausts much of the potential mass to be created from a companion, and any additional accretion might not allow for sufficient growth towards another explosion. The WD-kicks may also change the later evolution of the remnant binary. These, however, do not preclude a merger of the companion donor star remnant, once it becomes a WD with the current WD accretor, which then potentially follows a double-degenerate channel for type Ia SN, i.e. the same system could produce more than one SN, one for an SD channel and later from a DD channel.

The sharp transition in explosion outcome that we observe near $1.370~M_{\odot}$ provides important physical insight into the deflagration-to-detonation transition (DDT) process. Our simulations suggest that at this critical mass threshold, the central density reaches values where the deflagration burning becomes sufficiently energetic and the nuclear timescales become competitive with the hydrodynamic expansion timescales, allowing a transition to detonation. The existence of this threshold explains why SNe Iax and normal SNe Ia appear as distinct populations rather than forming a continuous spectrum of explosion properties. The threshold behavior creates a natural bifurcation in outcomes: WDs below the threshold produce SNe Iax, while those above it yield normal SNe Ia. However, this threshold is not solely determined by mass. As our results show, ignition geometry also plays a crucial role, particularly at masses near the transition point. The strong dependence on ignition conditions at masses around $1.365-1.370~M_{\odot}$ introduces an element of stochasticity in explosion outcomes, which could explain some of the observed diversity within both the SNe Iax and normal SNe Ia populations.

Our model makes several testable predictions for future observations. First, if most single-degenerate WDs undergo premature ignition as we propose, then SNe Iax should occur at rates comparable to or higher than normal SNe Ia from this channel. Theoretical rate estimates for the SD channel typically suggest $10-15\%$ of the current observational inferred type Ia rates. Current observational estimates suggest that SNe Iax represent approximately $5-30\%$ of the total SN Ia rate \cite{Foley2013, Miller2017}, which is consistent with our scenario where type Iax SNe are effectively the outcomes of the SD channel rather than normal Ia SNe. Second, our model predicts a strong positive correlation between ejecta mass, nickel yield, and expansion velocity within the SNe Iax class, driven primarily by variations in the progenitor WD mass. These are generally consistent with current observations [see also \cite{Foley2013}]. Future high-cadence surveys with instruments like the Vera C. Rubin Observatory will enable more systematic characterization of SNe Iax properties, allowing this prediction to be tested across a larger sample. Third, we support previous predictions \cite{Jordan2012b} of the existence of a population of kicked WD remnants with unusual compositions and thermal properties. These objects would appear as peculiar WDs with atmospheres contaminated by deflagration ash, moderately high velocities ($100-500$ km s$^{-1}$), and potentially inflated radii. Dedicated searches for such objects in our galactic neighborhood could provide direct confirmation of the partial deflagration scenario. In particular, current observations of peculiar hypervelocity WDs at hundreds of km s$^{-1}$ [e.g., \cite{ven+17}], may have already identified such remnants. Finally, if premature ignition is indeed common, we would expect a relative scarcity of super-Chandrasekhar or near-Chandrasekhar mass CO WDs in accreting binary systems, as these objects would be vulnerable to ignition before reaching such high masses. Demographic studies of WD masses in close binaries could test this prediction.

% ==================================
\section{Summary and Conclusions}
\label{sec:summary}
% ==================================
Our three-dimensional hydrodynamic simulations reveal that the ignition of sub-Chandrasekhar mass carbon-oxygen white dwarfs (CO WDs) in single-degenerate systems naturally produces partial deflagrations with properties that closely match observed Type Iax supernovae. These simulations demonstrate a sharp transition in explosion outcomes at approximately $1.370~M_{\odot}$: lower-mass WDs consistently undergo incomplete deflagrations that leave behind remnants, while higher-mass WDs experience complete disruption characteristic of normal Type Ia supernovae.

The partial deflagration outcomes reproduce key observational signatures of Type Iax events, including:
\begin{itemize}
    \item Moderate ejecta velocities ($3,000-5,000$ km s$^{-1}$)
    \item Modest $^{56}\text{Ni}$ yields ($0.03-0.1~M_{\odot}$)
    \item Substantial bound remnants ($0.9-1.2~M_{\odot}$)
    \item Asymmetric ejecta distributions
\end{itemize}
Our results suggest that the intrinsic diversity observed among Type Iax supernovae likely stems from variations in both WD mass and ignition conditions, with more offset ignitions generally producing weaker explosions compared to central ignitions at the same WD mass.

Critically, our findings indicate that premature ignition represents a fundamental limiting mechanism in the evolution of accreting WDs, analogous to but distinct from the constraints imposed by classical nova eruptions. This "safety valve" mechanism effectively interrupts the growth of WDs toward the Chandrasekhar limit, as WDs reaching masses of $1.32-1.37~M_{\odot}$ become increasingly susceptible to carbon ignition that results in partial deflagrations rather than complete detonations.

This has profound implications for supernova progenitor theory: the single-degenerate channel may predominantly produce Type Iax supernovae rather than normal Type Ia events. This constraint helps explain several observational puzzles, including the relative rates of Type Iax and Type Ia supernovae, and suggests that alternative progenitor scenarios—such as double-degenerate mergers or sub-Chandrasekhar double detonations—may be required to account for normal Type Ia supernovae.

Our model makes several testable predictions for future observations:
\begin{enumerate}
    \item A strong positive correlation between ejecta mass, nickel yield, and expansion velocity within the Type Iax class.
    \item The existence of a population of kicked WD remnants with unusual compositions, moderate velocities ($100-500$ km s$^{-1}$), and potentially inflated radii.
    \item A relative scarcity of super-Chandrasekhar or near-Chandrasekhar mass WDs in accreting binary systems.
\end{enumerate}
The mass threshold we identify (approximately $1.370~M_{\odot}$) provides important physical insight into the deflagration-to-detonation transition process and explains why Type Iax and normal Type Ia supernovae appear as distinct populations rather than forming a continuous spectrum of explosion properties.

Future work should focus on refining our understanding of ignition conditions in accreting WDs, exploring a wider range of ignition geometries and compositions, and developing more sophisticated nucleosynthesis calculations to enable direct comparison with observed spectra. Additionally, incorporating these findings into binary population synthesis models will be crucial for predicting the relative rates of different thermonuclear transients and assessing the overall contribution of the single-degenerate channel to the supernova population.

Our results significantly constrain the viability of the single-degenerate channel as a pathway toward classical Type Ia explosions, suggesting that the diversity of thermonuclear supernovae reflects not only variations in explosion mechanisms but also fundamental differences in progenitor evolution. This framework provides a coherent explanation for the observed diversity of thermonuclear transients and highlights the critical role of understanding sub-Chandrasekhar ignition processes in the broader context of stellar evolution and galactic chemical enrichment.

Finally, because premature ignition at sub-Chandrasekhar mass \emph{truncates} the high-mass accretion phase, it naturally alleviates the classical SSS and recurrent-nova tensions for SD progenitors (Methods).
Furthermore, in redirecting single-degenerate evolution toward Iax rather than normal Ia, premature sub-Chandrasekhar ignition also reconciles independent constraints on stripped H/He, surviving companions and early shocks, and radio/X-ray mass-loss limits, providing a unified resolution of the classic SD tensions.

% ==================================
\section{Methods}
\label{sec:methods}
% ==================================

% ==================================
\subsection{Hydrodynamic Simulations}
\label{sec:hydrodynamic_simulations}
% ==================================
We conducted comprehensive three-dimensional hydrodynamic simulations using the adaptive-mesh refinement (AMR) code FLASH version 4.7 \cite{Fryxell2000}. FLASH is particularly well-suited for modeling thermonuclear explosions due to its modular architecture, robust AMR capabilities, and specialized physics modules for nuclear burning and degenerate matter.

For all simulations, we employed the directionally unsplit hydrodynamic solver \cite{Lee2006, Lee2009, Lee2013}, which accurately captures three-dimensional shock propagation and complex flow patterns essential for modeling deflagration fronts. The degenerate matter was described using the Helmholtz equation of state \cite{Aparicio1998, Timmes1999, Timmes2000}, which incorporates electron degeneracy pressure, Coulomb corrections, radiation pressure, and thermal contributions from ions—all critical components for accurately modeling WD interiors across a wide range of densities and temperatures.

To simulate nuclear burning processes, we implemented the reaction-advection-diffusion flame model developed by Calder et al. \cite{Calder2007} and Townsley et al. \cite{Townsley2007, Townsley2009, Townsley2016}\footnote{\url{https://pages.astronomy.ua.edu/townsley/code/}}. This model tracks the propagation of the thermonuclear flame front through the degenerate material using an advection-diffusion-reaction equation approach, capturing both the subsonic deflagration phase and potential transitions to detonation. The nuclear reaction network follows the major energy-producing reactions including carbon, oxygen, and helium burning channels, accurately accounting for the energy release and nuclear statistical equilibrium (NSE) conditions reached in the densest regions.
\\
\\

% ==================================
\subsection{Initial Models and Simulation Setup}
\label{sec:init_and_setup}
% ==================================
We constructed a series of hydrostatic CO WD models with masses ranging from $1.325$ to $1.374~M_{\odot}$, corresponding to central densities between $4.8 \times 10^8$ and $1.2 \times 10^9$ g cm$^{-3}$. Each initial model was generated by numerically integrating the hydrostatic equilibrium equation using the Helmholtz EOS with a uniform initial temperature of $10^7$ K and a uniform 50/50 carbon-oxygen composition by mass.

These WD models were mapped onto the three-dimensional FLASH grid and allowed to relax over several dynamical timescales to ensure hydrostatic equilibrium before ignition. Our computational domain spanned $(1 \times 10^{11}\text{ cm})^3$ with 15 levels of refinement, providing an effective resolution of $7.63$ km at the highest refinement level. This resolution is sufficient to capture the propagation and development of the deflagration fronts while remaining computationally feasible for a systematic parameter study.

\begin{figure}
\includegraphics[width=0.5\columnwidth]{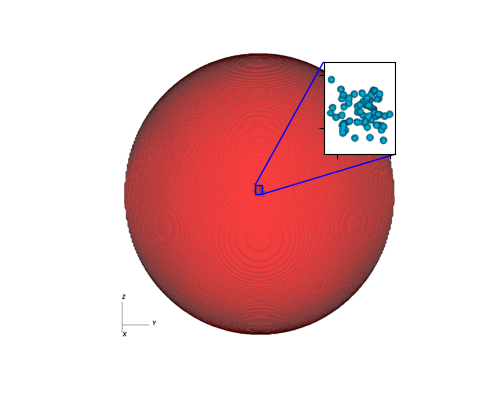}
\caption{\textbf{Initial Ignition Setup.} The red sphere represents the white dwarf (WD). The inner region indicates the initial deflagration zone. A zoomed-in view of this region is shown in the upper right corner. The blue spheres represent the ignited ("ash") points. These are confined within a sphere of 128 km radius, centered at either $z = 88$ km or $z = 168$ km. Each point has a radius of 16 km, and there are a total of 63 such points spread randomly inside the confined region.}
\label{fig:init-ignition}
\end{figure}

To initiate the deflagration process, we used a stochastic multi-point ignition scenario. Specifically, we positioned 63 ignition points randomly within a spherical shell of radius 128 km, centered at two different offset positions along the $z$-axis: $z=88$ km and $z=168$ km from the WD center. Each ignition kernel was modeled as a spherical region of radius 16 km with fully burned material ("ash") at an elevated temperature corresponding to the energy release from carbon fusion (Figure \ref{fig:init-ignition}). This multi-point ignition setup allows us to systematically investigate how the asymmetry and geometry of the initial burning region affect the overall explosion dynamics.

Table~\ref{tab:models} summarizes the key parameters for all simulation models in our study, including the WD masses, central densities, ignition configurations, and resulting explosion properties. For each model, we tracked the simulation until either the deflagration fully consumed the WD or the partial burning phase reached a quasi-steady state (typically 2-3 seconds of physical time), after which we analyzed the ejecta properties, nucleosynthetic yields, and remnant characteristics.

% ==================================
\subsection{Analysis Methods}
\label{sec:analysis_methods}
% ==================================
To analyze the simulation results, we tracked multiple explosion metrics throughout each simulation. These included the mass of synthesized radioactive $^{56}\text{Ni}$, total ejected mass, remnant mass (if any), characteristic ejecta velocities, and explosion energy. The ejecta was defined as material with positive total energy (kinetic plus potential plus internal), while bound material was considered part of the remnant.

For each model, we calculated the mass-weighted mean radial velocity of the ejecta to characterize the typical expansion speeds. We also tracked the morphology of the deflagration front to identify whether the burning remained asymmetric (leading to partial deflagration) or encompassed the entire star (leading to complete disruption). The asymmetry of the ejecta was quantified using the center-of-mass offset from the initial WD center, allowing us to estimate the resulting kick velocities of any remnants. We calculate the nucleosynthesis in post-processing using a 136-isotope network and between 350,000 and 500,000 Lagrangian particles.

We performed these simulations on the Technion Nyx high-performance computer using 576 CPU cores per simulation, with typical run times of approximately 414,720 CPU hours per model.

% ==================================
\subsection{Impact of sub-Chandrasekhar ignition on SSS exposure and high-mass recurrent novae}
\label{sec:SSS-RNe}
% ==================================
\paragraph{Aim.}
We provide a transparent, closed-form estimate of how lowering the effective ignition mass $M_{\rm crit}$ (relative to a reference $1.38\,M_\odot$) reduces (i) the time accreting WDs spend in the steady H-burning (SSS) regime and (ii) the contribution of high-mass recurrent novae (RNe) along SD evolutionary tracks. The intent is to isolate the lever arm supplied by truncating the high-mass phase; full population synthesis is beyond scope.

\paragraph{Definitions and assumptions.}
A WD grows from $M_0$ to $M_{\rm crit}$ with mass-transfer rate $\dot{M}$. The track first enters stable H-burning at $M_{\rm s}$ (thus SSS for $M\in[M_{\rm s},M_{\rm crit}]$). The subset of high-mass RNe occupies a neighboring unstable regime $M\in[M_{\rm rn},M_{\rm crit}]$, with $M_{\rm rn}\lesssim M_{\rm s}$ (systems just below stability at high $M$). Over the narrow interval of interest ($M\!\approx\!1.30$–$1.38\,M_\odot$), the $M$-dependences of the stable-strip retention and effective accretion rate are weak; for RNe we adopt the standard nova ignition scaling $M_{\rm ign}\propto M^{-\alpha}\dot{M}^{-\beta}$ with $\alpha\simeq 3$–$4$, $\beta\simeq 0.5$–$0.7$ (see reviews cited in the main text, e.g., \cite{Maoz2014review}).

\paragraph{SSS exposure.}
For one accretion track, the SSS time scales as
\begin{equation}
T_{\rm SSS}(M_{\rm crit}) \;\propto\; \int_{M_{\rm s}}^{M_{\rm crit}} \frac{{\rm d}M}{\eta_{\rm st}(M)\,\dot{M}_{\rm st}(M)} ,
\label{eq:S1}
\end{equation}
where $\eta_{\rm st}$ is the net (H$\times$He) retention in the stable strip and $\dot{M}_{\rm st}$ the accretion rate while stable. Linearizing the product $\eta_{\rm st}\dot{M}_{\rm st}$ across $1.30$–$1.38\,M_\odot$ yields the ratio
\begin{equation}
    \begin{array}{r}
\frac{T_{\rm SSS}(M_{\rm crit})}{T_{\rm SSS}(1.38)} \;\approx\; 
\frac{M_{\rm crit}-M_{\rm s}}{1.38-M_{\rm s}} \;\times\; \mathcal{W}_{\rm SSS}, \\
\qquad \mathcal{W}_{\rm SSS}\in[0.9,1.1],
    \end{array}
\label{eq:S2}
\end{equation}
with $\mathcal{W}_{\rm SSS}$ absorbing the mild $M$-dependences. The fraction of SSS exposure \emph{removed} by lowering the cap from $1.38$ to $M_{\rm crit}^{\rm new}$ is
\begin{equation}
\begin{array}{rl}    
    f_{\rm SSS}^{\rm removed} \; & \equiv\; 1 - \frac{T_{\rm SSS}(M_{\rm crit}^{\rm new})}{T_{\rm SSS}(1.38)} \\
    \;& \approx\; 1 - \frac{M_{\rm crit}^{\rm new}-M_{\rm s}}{1.38-M_{\rm s}} \;\;(\times\,\mathcal{W}_{\rm SSS}^{-1}).
\end{array}
\label{eq:S3}
\end{equation}

\paragraph{High-mass recurrent novae.}
In the unstable regime, cycles accumulate at rate $\dot{M}/M_{\rm ign}$. A continuum estimate for the number of RN cycles accrued while the WD \emph{grows} by ${\rm d}M$ gives
\begin{equation}
\begin{array}{rl}
    {\rm d}N_{\rm RN} & \;\simeq\; \frac{\dot{M}}{M_{\rm ign}(M,\dot{M})}\,\frac{{\rm d}M}{\eta_{\rm un}(M)\,\dot{M}} \\
    & \,=\, \frac{{\rm d}M}{\eta_{\rm un}(M)\,M_{\rm ign}(M)} \;\propto\; M^{\alpha}{\rm d}M,
\end{array}
\label{eq:S4}
\end{equation}
hence, neglecting the slow variation of $\eta_{\rm un}$ across $1.2$–$1.38\,M_\odot$,
\begin{equation}
N_{\rm RN}(M_{\rm crit}) \;\propto\; \int_{M_{\rm rn}}^{M_{\rm crit}} M^{\alpha}\,{\rm d}M 
\,=\, \frac{M_{\rm crit}^{\alpha+1}-M_{\rm rn}^{\alpha+1}}{\alpha+1}.
\label{eq:S5}
\end{equation}
The fraction of high-mass RN cycles \emph{removed} when lowering the cap from $1.38$ to $M_{\rm crit}^{\rm new}$ is therefore
\begin{equation}
f_{\rm RN}^{\rm removed} \;=\; 
\frac{1.38^{\alpha+1} - \left(M_{\rm crit}^{\rm new}\right)^{\alpha+1}}
{1.38^{\alpha+1} - M_{\rm rn}^{\alpha+1}} .
\label{eq:S6}
\end{equation}

\paragraph{Worked values.}
To provide concrete numbers without hidden code, we adopt $M_{\rm s}=1.30\,M_\odot$, $\alpha=3.5$, and two plausible onsets for the high-mass RN subset, $M_{\rm rn}=1.20$ and $1.25\,M_\odot$. With $\mathcal{W}_{\rm SSS}=1$, Eq.~(\ref{eq:S3}) gives
\[
f_{\rm SSS}^{\rm removed}
= 1-\frac{M_{\rm crit}^{\rm new}-1.30}{1.38-1.30},
\]
so that
\[
\begin{array}{lcl}
M_{\rm crit}^{\rm new}=1.35: & f_{\rm SSS}^{\rm removed}=0.375 \;(37.5\%),\\
M_{\rm crit}^{\rm new}=1.32: & f_{\rm SSS}^{\rm removed}=0.75 \;(75\%),\\
M_{\rm crit}^{\rm new}=1.30: & f_{\rm SSS}^{\rm removed}=1.00 \;(100\%).
\end{array}
\]
For RNe, with $\alpha+1=4.5$ and
\[\begin{array}{ccc} 
1.38^{4.5}\!\approx\!4.28,\;\; &
1.35^{4.5}\!\approx\!3.94,\;\; &
1.32^{4.5}\!\approx\!3.61,\;\; \\
1.30^{4.5}\!\approx\!3.36,\;\; &
1.25^{4.5}\!\approx\!2.74,\;\; &
1.20^{4.5}\!\approx\!2.21,
\end{array}
\]
Eq.~(\ref{eq:S6}) yields the fractions listed in Table~\ref{tab:S1}.

\paragraph{Sensitivity and reproducibility.}
Varying $\mathcal{W}_{\rm SSS}$ over $[0.9,1.1]$ perturbs the SSS percentages by $\pm 10\%$ of their values. Adopting $\alpha\in[3.0,4.0]$ shifts the RN percentages in Table~\ref{tab:S1} by $\lesssim 5$–$10$ points.  Empirical context for SSS/RN constraints is reviewed in \cite{Maoz2014review}.

% TTTTTTTTTTTTTTTTTTTTTTTTTTTTTTTTTTTTTTTTTT
% Table {tab:S1}
% TTTTTTTTTTTTTTTTTTTTTTTTTTTTTTTTTTTTTTTTTT

% % ==================================
% \section*{Data Availability}
% \label{sec:data_availability}
% % ==================================
% The datasets generated and analyzed during the current study will be made available upon reasonable request to the corresponding author.

% ==================================
% \section*{Code Availability}
% \label{sec:code_availability}
% % ==================================
% The FLASH code is an open-source, publicly available multiphysics simulation code (\url{https://flash.uchicago.edu/}). Specific modifications and input files used for these simulations can be made available upon reasonable request.

% ======================================
\section*{Acknowledgements}
\label{sec:acknowledge}
% ======================================
We acknowledge support for this project from the European Union's Horizon 2020 research and innovation program under grant agreement No 865932-ERC-SNeX.

% =================================
% =================================

%%% Below is for using the bib file
\bibliography{snIa}
\bibliographystyle{aasjournal}
%\bibliographystyle{mnras}

% TTTTTTTTTTTTTTTTTTTTTTTTTTTTTTTTTTTTTTTTTT
\begin{table*}
\centering
\caption{\textbf{Fraction of SSS exposure and high-mass RN cycles removed when lowering the cap from $1.38\,M_\odot$.}
SSS assumes $M_{\rm s}=1.30\,M_\odot$; RNe use $\alpha=3.5$ with two values of $M_{\rm rn}$.}
\label{tab:S1}
\begin{tabular}{lccc}
\hline
$M_{\rm crit}^{\rm new}$ & SSS removed & RN removed ($M_{\rm rn}=1.20$) & RN removed ($M_{\rm rn}=1.25$) \\
\hline
$1.35\,M_\odot$ & $37.5\%$ & $16\%$ & $22\%$ \\
$1.32\,M_\odot$ & $75\%$   & $32\%$ & $43\%$ \\
$1.30\,M_\odot$ & $100\%$  & $44\%$ & $60\%$ \\
\hline
\end{tabular}
\end{table*}
% TTTTTTTTTTTTTTTTTTTTTTTTTTTTTTTTTTTTTTTTTT

% TTTTTTTTTTTTTTTTTTTTTTTTTTTTTTTTTTTTTTTTTT
\begin{table*}
\centering
\caption{\textbf{Summary of Simulation Parameters and Outcomes.} The columns show: model designation, ignition offset along the $z$-axis (km), WD mass ($M_{\odot}$), central density ($10^9$ g cm$^{-3}$), ejecta mass ($M_{\odot}$), characteristic ejecta velocity (km s$^{-1}$), and synthesized $^{56}\text{Ni}$ mass ($M_{\odot}$). Models w1374-z88 and w1370-z168 are the only models to achieve full detonation and no bounded mass remains.}
\label{tab:models}
\begin{tabular}{lcccccc}
\hline
Model name & Offset (km) & Mass ($M_{\odot}$) & $\rho_c$ ($10^9$ g cm$^{-3}$) & M$_{\text{ej}}$ ($M_{\odot}$) & V$_{\text{ej}}$ (km s$^{-1}$) & $^{56}\text{Ni}$ ($M_{\odot}$) \\
\hline
w1325-z88 & 88 & 1.325 & 0.488 & 0.428 & 4410 & 0.041 \\
w1330-z88 & 88 & 1.330 & 0.518 & 0.389 & 4499 & 0.053 \\
w1335-z88 & 88 & 1.335 & 0.568 & 0.408 & 4527 & 0.058 \\
w1340-z88 & 88 & 1.340 & 0.608 & 0.417 & 4559 & 0.062 \\
w1350-z88 & 88 & 1.350 & 0.728 & 0.480 & 4777 & 0.072 \\
w1355-z88 & 88 & 1.355 & 0.788 & 0.453 & 4717 & 0.073 \\
w1360-z88 & 88 & 1.360 & 0.868 & 0.469 & 4764 & 0.079 \\
w1365-z88 & 88 & 1.365 & 0.958 & 0.484 & 4795 & 0.090 \\
w1370-z88 & 88 & 1.370 & 1.190 & 0.487 & 4871 & 0.108 \\
w1374-z88 & 88 & 1.374 & 1.290 & 1.364 & 9751 & 0.893 \\
\hline
w1330-z168 & 168 & 1.330 & 0.518 & 0.085 & 3409 & 0.025 \\
w1350-z168 & 168 & 1.350 & 0.728 & 0.113 & 3483 & 0.040 \\
w1365-z168 & 168 & 1.365 & 0.958 & 0.105 & 3653 & 0.051 \\
w1370-z168 & 168 & 1.370 & 1.190 & 1.367 & 8933 & 0.593 \\
\hline
\end{tabular}
\end{table*}
% TTTTTTTTTTTTTTTTTTTTTTTTTTTTTTTTTTTTTTTTTT

% FFFFFFFFFFFFFFFFFFFFFFFFFFFFFFFFFFFFFFFFFF
% FFFFFFFFFFFFFFFFFFFFFFFFFFFFFFFFFFFFFFFFFF

% FFFFFFFFFFFFFFFFFFFFFFFFFFFFFFFFFFFFFFFFFF
% FFFFFFFFFFFFFFFFFFFFFFFFFFFFFFFFFFFFFFFFFF

% FFFFFFFFFFFFFFFFFFFFFFFFFFFFFFFFFFFFFFFFFF
% FFFFFFFFFFFFFFFFFFFFFFFFFFFFFFFFFFFFFFFFFF

% FFFFFFFFFFFFFFFFFFFFFFFFFFFFFFFFFFFFFFFFFF
% FFFFFFFFFFFFFFFFFFFFFFFFFFFFFFFFFFFFFFFFFF

% FFFFFFFFFFFFFFFFFFFFFFFFFFFFFFFFFFFFFFFFFF
% FFFFFFFFFFFFFFFFFFFFFFFFFFFFFFFFFFFFFFFFFF

% FFFFFFFFFFFFFFFFFFFFFFFFFFFFFFFFFFFFFFFFFF
% FFFFFFFFFFFFFFFFFFFFFFFFFFFFFFFFFFFFFFFFFF

% FFFFFFFFFFFFFFFFFFFFFFFFFFFFFFFFFFFFFFFFFF
% FFFFFFFFFFFFFFFFFFFFFFFFFFFFFFFFFFFFFFFFFF

% FFFFFFFFFFFFFFFFFFFFFFFFFFFFFFFFFFFFFFFFFF
% FFFFFFFFFFFFFFFFFFFFFFFFFFFFFFFFFFFFFFFFFF

\end{document}